\documentstyle [12pt] {article}
\normalbaselines
\begin {document}
\begin {center}
{\bf METHOD OF REPLACING THE VARIABLES FOR GENERALIZED
SYMMETRY OF D'ALEMBERT EQUATION} \\ [3mm]
{G.A. Kotel'nikov} \\
{\small RRC Kurchatov Institute, Kurchatov Sq. 1, Moscow 123182,
Russia} \\ 
E-mail: kga@electronics.kiae.ru \\ [4mm]
\parbox{14.0cm}{\small It is shown that by  generalized  understanding
of  symmetry  the  D'Alembert  equation  for  one  component  field is
invariant   with   respect   to   arbitrary   reversible    coordinate
transformations.}
\bigskip \end{center}
\begin{sloppypar}\noindent
Symmetries play  an  important  role  in  particle physics and quantum
field theory [1], nuclear physics [2], mathematical physics [3]. It is
proposed some receptions for finding the symmetries,  for example, the
method of  replacing  the  variables [4],  the Lie algorithm [3],  the
theoretical-algebraic approach [5]. The purpose of the present work is
the generalization of the method of replacing the variables.  We start
from the following Definition of symmetry. \end{sloppypar}

\newtheorem{definition}{Definition}
\label{d1}
\begin{definition} Let  some  partial  differential   equation   $\hat
L'\phi'(x')=0$ be given.  By symmetry of this equation with respect to
the variables replacement $x'=x'(x)$, $\phi'=\phi'(\Phi\phi)$ we shall
understand  the  compatibility  of the engaging equations system $\hat
A\phi'(\Phi\phi)=0$,     $\hat      L\phi(x)=0$,      where      $\hat
A\phi'(\Phi\phi)=0$ is obtained from the initial equation by replacing
the variables, $\hat L'=\hat L$, $\Phi(x)$ is some weight function. If
the  equation  $\hat  A\phi'(\Phi\phi)=0$  can be transformed into the
form $\hat L(\Psi\phi)=0$, the symmetry will be named the standard Lie
symmetry, otherwise the generalized symmetry.
\end{definition}
Elements of  this  Definition were used to study the Maxwell equations
symmetries [6-8].  In the present work we shall apply Definition 1 for
investigation of symmetries of the one-component D'Alembert equation:
\begin{equation}
\label{f1}
\Box'\phi'(x')=
\frac{\partial^2 \phi'}{\partial x_1'^2}+
\frac{\partial^2 \phi'}{\partial x_2'^2}+
\frac{\partial^2 \phi'}{\partial x_3'^2}+
\frac{\partial^2 \phi'}{\partial x_4'^2}=0.
\end{equation}
Let us  introduce  the arbitrary reversible coordinate transformations
$x'=x'(x)$   and   the   transformation   of   the   field    variable
$\phi'=\phi(\Phi\phi)$,  where $\Phi(x)$ is some unknown function,  as
well     as     take     into     account      $\partial\phi'/\partial
x_i'=\sum_{j}(\partial\phi'    /\partial\xi)(\partial\Phi\phi/\partial
x_j)(\partial x_j/\partial x_i')$, $\partial^2\phi'/\partial {x_i'}^2=
\sum_{j}(\partial^2x_j/\partial   {x_i'}^2)(\partial\phi'/\partial\xi)
(\partial\Phi\phi/\partial                   x_j)                    +
\sum_{jk}(\partial^2\Phi\phi/\partial   x_j\partial   x_k)   (\partial
x_j/\partial      x_i')      (\partial      x_k/      \\      \partial
x_i')(\partial\phi'/\partial\xi)      +     \sum_{jk}(\partial^2\phi'/
\partial\xi^2)                              (\partial\Phi\phi/\partial
x_j)(\partial\Phi\phi/\partial    x_k)(\partial   x_j/\partial   x_i')
(\partial x_k/\partial x_i')$,  where $\xi=\Phi\phi$.  After replacing
the variables we find that the equation $\Box'\phi'=0$ transforms into
itself,  if  the  system  of  the  engaging  equations  is   fulfilled
\begin{equation}     \label{f2}     \begin{array}{l}     \displaystyle
\sum_{i}\sum_{j}\frac{\partial^2        x_j}{\partial        {x_i'}^2}
\frac{\partial\phi'}{\partial\xi}     \frac{\partial\Phi\phi}{\partial
x_j}+      \sum_{i}\sum_{j=k}\biggl(\frac{\partial       x_j}{\partial
x_i'}\biggr)^2                       \frac{\partial\phi'}{\partial\xi}
\frac{\partial^2\Phi\phi}{\partial                           {x_j}^2}+
\sum_{i}\sum_{j<k}\sum_{k}2\frac{\partial      x_j}{\partial     x_i'}
\frac{\partial x_k}{\partial x_i'} \frac{\partial  \phi'}{\partial\xi}
\frac{\partial^2\Phi\phi}{\partial  x_j\partial  x_k}+ \\ \vspace{3mm}
\displaystyle  \sum_{i}\sum_{j=k}\biggl(\frac{\partial   x_j}{\partial
x_i}\biggr)^2          \frac{\partial^2          \phi'}{\partial\xi^2}
\biggl(\frac{\partial\Phi\phi}{\partial  x_j}\biggr)^2+  \displaystyle
\sum_{i}\sum_{j<k}\sum_{k}2\frac{\partial      x_j}{\partial     x_i'}
\frac{\partial                   x_k}{\partial                   x_i'}
\frac{\partial^2\phi'}{\partial\xi^2} \frac{\partial\Phi\phi}{\partial
x_j}  \frac{\partial\Phi\phi}{\partial   x_k}   =0;   \\   \Box\phi=0.
\end{array} \end{equation} Here $x=(x_1,x_2,x_3,x_4),  \ x_4=ict$, $c$
is the speed of light,  $t$ is the time.  Let us put the  solution  of
D'Alembert  equation  $\phi$  into  the  first  equation  of  the  set
(\ref{f2}).  If the obtained equation has a  solution,  then  the  set
(\ref{f2}) will be compatible. According to Definition 1 it will mean
that  the  arbitrary  reversible  transformations  $x'=x'(x)$  are the
symmetry transformations of the initial equation $\Box'\phi'=0$. Owing
to presence of the expressions $(\partial\Phi\phi/\partial x_j)^2$ and
$(\partial\Phi\phi/\partial  x_j)(\partial\Phi\phi/\partial  x_k)$  in
the first equation from the set (\ref{f2}),  the latter has non-linear
character.  Since the analysis of non-linear systems is  difficult  we
suppose     that    \begin{equation}    \label{f3}    \begin{array}{l}
\displaystyle   \frac{\partial^2\phi'}{\partial\xi^2}=0.   \end{array}
\end{equation}  In  this  case  the  non-linear  components in the set
(\ref{f2}) turn to zero and the system will be linear.  As  result  we
find   the  field  transformation  law  by  integrating  the  equation
(\ref{f3})
\begin{equation} \label{f4}        \phi'={\rm        C}_1\Phi\phi+{\rm
C}_2\to\phi'=\Phi\phi.  \end{equation} Here we suppose for  simplicity
that the constants of integration are ${\rm C}_1=1,  {\rm C}_2=0$.  It
is this law of field transformation that was used within the algorithm
[7,  8].  It  marks  the  position of the algorithm in the generalized
variables  replacement  method.  Taking  into  account  the   formulae
(\ref{f3})  and (\ref{f4}),  we find the following form for the system
(\ref{f2}):     \begin{equation}     \label{f5}      \begin{array}{ll}
\displaystyle        \frac{\partial^2\phi'}{\partial\xi^2}=0;        &
\phi'=\Phi\phi;  \\  \vspace{2mm}  \displaystyle   \sum_{j}\Box'   x_j
\frac{\partial\Phi\phi}{\partial                                 x_j}+
\sum_{i}\sum_{j}\biggl(\frac{\partial   x_j}{\partial   x_i'}\biggr)^2
\frac{\partial^2\Phi\phi}{\partial                           {x_j}^2}+
\sum_{i}\sum_{j<k}\sum_{k}2\frac{\partial     x_j}{\partial      x_i'}
\frac{\partial  x_k}{\partial x_i'} \frac{\partial^2\Phi\phi}{\partial
x_j\partial x_k}=0;  & {} \\  \Box\phi=0.  \end{array}  \end{equation}
Since  here  $\Phi(x)=\phi'(x'\to  x)/\phi(x)$,  where $\phi'(x')$ and
$\phi(x)$  are  the  solutions  of  D'Alembert  equation,  the  system
(\ref{f5}) has a common solution and consequently is compatible.  This
means that the arbitrary  reversible  transformations  of  coordinates
$x'=x'(x)$   are   symmetry   transformations  for  the  one-component
D'Alembert equation if the field  is  transformed  with  the  help  of
weight function $\Phi(x)$ according to the law (\ref{f4}). The form of
this function depends on D'Alembert equation solutions and the law  of
the coordinate transformations $x'=x'(x)$.

Next we shall consider the following examples.

Let the  coordinate  transformations  belong  to  the  {\it Poincar\'e
group} $P_{10}$:  \begin{equation} \label{f6} x_j'={\it  L}_{jk}x_k  +
a_j,  \end{equation} where $L_{jk}$ is the matrix of the Lorentz group
$L_6$,  $a_j$ are the parameters of the translation  group  $T_4$.  In
this    case    we   have   $\Box'x_j=\sum_k   L_{jk}'\Box'   x'_k=0$,
$\sum_{i}(\partial  x_j/\partial  x_i')(\partial  x_k/\partial  x_i')=
\sum_{i}  L_{ji}'L_{ki}'=\delta_{jk}$.  The  last  term  in the second
equation (\ref{f5}) turns  to  zero.  The  set  reduces  to  the  form
\begin{equation}     \label{f7}    \Box\Phi\phi=0;    \    \Box\phi=0.
\end{equation} According to Definition1 1 this is a sign  of  the  Lie
symmetry.   The   weight   function   belongs   to  the  set  in  [8]:
\begin{equation}                                            \label{s1}
\Phi_{P_{10}}(x)=\frac{\phi'(x)}{\phi(x)}\in\biggl\{1;
\frac{1}{\phi(x)};                         \frac{P_j\phi(x)}{\phi(x)};
\frac{M_{jk}\phi(x)}{\phi(x)};          \frac{P_jP_k\phi(x)}{\phi(x)};
\frac{P_jM_{kl}\phi(x)}{\phi(x)};  \cdots\biggr\} \end{equation} where
$P_j,   \   M_{jk}$   are   the   generators   of   Poincar\'e  group,
$j,k,l=1,2,3,4$. In the space of D'Alembert equation solutions the set
defines  a rule of the change from a solution to solution.  The weight
function      $\Phi(x)=1\in\Phi_{P_{10}}(x)$      determines       the
transformational properties of the solutions $\phi'=\phi$, which means
the well-known relativistic symmetry of D'Alembert equation [9, 10].

Let the transformations of coordinates belong to the {\it Weyl  group}
$W_{11}$:  \begin{equation} \label{g7} x_j'=\rho{\it L}_{jk}x_k + a_j,
\end{equation} where  $\rho$=const  is  the  parameter  of  the  scale
transformations  of  the  group  $\Delta_1$.  In  this  case  we  have
$\Box'x_j=\rho'\sum_k    L_{jk}'\Box'    x'_k=0$,     $\sum_i(\partial
x_j/\partial    x_i')(\partial   x_k/\partial   x_i')=   \sum_i\rho'^2
L_{ji}'L_{ki}'=\rho'^2\delta_{jk}=\rho^{-2}\delta_{jk}$.    The    set
(\ref{f5})  reduces  to  the  set  (\ref{f7})  and  has  the  solution
$\Phi_{W_{11}}={\rm  C}\Phi_{P_{10}}$,  where  ${\rm  C}$=const.   The
weight  function  $\Phi(x)={\rm  C}$  and  the law $\phi'={\rm C}\phi$
means the well-known Weyl symmetry of D'Alembert equation [9, 10]. Let
here ${\rm C}$ be equal $\rho^l$, where $l$ is the conformal dimension
\footnote{The conformal dimension is the number $l$ characterizing the
behavior  of  the  field   under  scale transformations $x'=\rho x,  \
\phi'(x')=\rho^l\phi(x)$ [11].} of the field $\phi(x)$.  Consequently,
D'Alembert  equation  is  $W_{11}$-invariant for the field $\phi$ with
arbitrary conformal dimension $l$.  This property is essential for the
Voigt [4] and Umov [12] works as will be shown just below.

Let the coordinate transformations belong to the {\it Inversion group}
$I$:  \begin{equation}  \label{ff8}  \begin{array}{lll}  \displaystyle
x_j'=-\frac{x_j}{x^2};    &   x^2={x_1}^2+{x_2}^2+{x_3}^2+{x_4}^2;   &
x^2x'^2=1.  \end{array}  \end{equation}   In   this   case   we   have
$\Box'x_j=4x'_j/x'^4=-4x_jx^2$,    \  $\sum_i(\partial    x_j/\partial
x'_i)(\partial                   x_k/\partial                   x_i')=
\rho'^2(x')\delta_{jk}=1/x'^4\delta_{jk}=x^4\delta_{jk}$.    The   set
(\ref{f5})  reduces   to   the   set:   \begin{equation}   \label{fg8}
-4x_j\frac{\partial   \Phi\phi}{\partial  x_j}+  x^2\Box\Phi\phi=0;  \
\Box\phi=0.  \end{equation} The substitution  of  $\Phi(x)=x^2\Psi(x)$
transforms  the  equation  (\ref{fg8}) for $\Phi(x)$ into the equation
$\Box\Psi\phi=0$ for $\Psi(x)$.  It is a sign of the Lie symmetry. The
equation  has  the  solution  $\Psi=1$.  The  result is $\Phi(x)=x^2$.
Consequently,   the   field   transforms   according   to   the    law
$\phi'=x^2\phi(x)=\rho^{-1}(x)\phi(x)$.   This   means  the  conformal
dimension $l=-1$ of the field $\phi(x)$  in  the  case  of  D'Alembert
equation symmetry with respect to the Inversion group $I$ in agreement
with [5,  10].  In a general case the weight function belongs  to  the
set:     \begin{equation}     \label{ss2}    \Phi_{I}(x)=x^2\Psi(x)\in
\biggl\{x^2;    \frac{x^2}{\phi(x)};    x^2\frac{P_j\phi(x)}{\phi(x)};
x^2\frac{M_{jk}\phi(x)}{\phi(x)};    x^2\frac{P_jP_k\phi(x)}{\phi(x)};
\cdots\biggr\}. \end{equation}

Let the  coordinate  transformations  belong  to  the   {\it   Special
Conformal      Group}     $C_{4}$:     \begin{equation}     \label{g8}
\begin{array}{lll} \displaystyle x_j'=\frac{x_j-a_jx^2}{\sigma(x)};  &
\sigma(x)=1-2a\cdot    x+a^2x^2;    &   \sigma\sigma'=1.   \end{array}
\end{equation} In this case we have $\Box'x_j=4(a_j-a^2x_j)\sigma(x)$,
\ $\sum_i(\partial  x_j/\partial  x_i')(\partial  x_k/\partial  x_i')=
\rho'^2(x')\delta_{jk}=\sigma^2(x)\delta_{jk}$.  The  set   (\ref{f5})
reduces    to    the    set:   \begin{equation}   \label{f8}   4\sigma
(x)(a_j-a^2x_j)\frac{\partial         \Phi\phi}{\partial         x_j}+
\sigma^2(x)\Box\Phi\phi=0;    \    \Box\phi=0.    \end{equation}   The
substitution of  $\Phi(x)=\sigma(x)\Psi(x)$  transforms  the  equation
(\ref{f8}) into the equation $\Box\Psi\phi=0$ which corresponds to the
Lie   symmetry.    From    this    equation    we    have    $\Psi=1$,
$\Phi(x)=\sigma(x)$.   Therefore   $\phi'=\sigma(x)\phi(x)$   and  the
conformal dimension of the field is $l=-1$ as  above.  Analogously  to
(\ref{ss2}),  the weight function belongs to the set: \begin{equation}
\label{s2}    \Phi_{C_4}(x)=\sigma(x)\Psi(x)\in     \biggl\{\sigma(x);
\frac{\sigma(x)}{\phi(x)};        \sigma(x)\frac{P_j\phi(x)}{\phi(x)};
\sigma(x)\frac{M_{jk}\phi(x)}{\phi(x)}; \cdots\biggr\}. \end{equation}
From  here  we  can  see that $\phi(x)=1/\sigma(x)$ is the solution of
D'Alembert equation. Combination of $W_{11}$, $I$ and $C_4$ symmetries
means  the  well-known D'Alembert equation conformal $C_{15}$-symmetry
[5, 9, 10].

Let the  coordinate  transformations belong to the {\it Galilei group}
$G_{1}$:  \begin{equation} \label{f9} x_1'=x_1+i\beta x_4; \ x_2'=x_2;
\  x_3'=x_3;  \ x_4'=\gamma x_4;  \ c'=\gamma c,  \end{equation} where
$\beta'=-\beta/\gamma$,  \  $\gamma'=1/\gamma$,   \   $\beta=V/c$,   \
$\gamma=(1-2\beta   n_x+\beta^2)^{1/2}$.   In   this   case   we  have
$\Box'x_j=0$,  \ $\sum_{i}(\partial x_1/\partial x_i')^2=1-\beta'^2; \
\sum_{i}(\partial  x_2/  \partial  x_i')^2  =  \sum_{i}(\partial  x_3/
\partial  x_i')^2=1;  \  \sum_{i}(\partial  x_4/\partial   x_i')^2   =
\gamma'^2;    \    \sum_{i}(\partial    x_1/\partial    x_i')(\partial
x_2/\partial x_i')  =  \sum_{i}(\partial  x_1/\partial  x_i')(\partial
x_3/\partial x_i') = \sum_{i}(\partial x_2/ \\ \partial x_i')(\partial
x_3/\partial x_i')  =  \sum_{i}(\partial  x_2/\partial  x_i')(\partial
x_4/\partial x_i')=0; \ \sum_{i}(\partial x_1/\partial x_i') (\partial
x_4/\partial x_i') = i\beta'\gamma'=-i\beta/\gamma^2$.  After  putting
these   expressions   into   the   set   (\ref{f5})   we   find   [8]:
\begin{equation}       \label{f10}       \Box\Phi\phi-\frac{\partial^2
\Phi\phi}{\partial    {x_4}^2}-    \biggl(i\frac{\partial   }{\partial
x_4}+\beta\frac{\partial       }        {\partial        x_1}\biggr)^2
\frac{\Phi\phi}{\gamma^2}=
\biggl[\frac{(i\partial_4+\beta\partial_1)^2}{\gamma^2}-
\triangle\biggr]\Phi\phi=0.    \end{equation}   In   accordance   with
Definition 1 it means that the Galilei symmetry of D'Alembert equation
is  the  generalized  symmetry  (being  the conditional one [8]).  The
weight function belongs to the set  [7]:  \begin{equation}  \label{s3}
\Phi_{G_1}(x)=\frac{\phi'(x'\to                           x)}{\phi(x)}
\in\biggl\{\frac{\phi(x')}{\phi(x)};                \frac{1}{\phi(x)};
\frac{P'_j\phi(x')}{\phi(x)};\frac{[\Box',H'_1]\phi(x')}
{\phi(x)};\cdots\biggr\}, \end{equation} where $H'_1=it'\partial_{x'}$
is the  generator  of the pure Galilei transformations.  For the plane
waves   the weight function  $\Phi(x)$  is  [6 - 8]:  \begin{equation}
\label{f12}        \Phi_{G_1}(x)=\frac{\phi(x'\to        x)}{\phi(x)}=
exp\biggl\{-\frac{i}{\gamma}\biggl[\biggl(1-\gamma\biggr)k       \cdot
x-\beta\omega\biggl(n_xt-\frac{x}{c}\biggr)\biggr]\biggr\},
\end{equation} where $k=({\bf k},k_4)$,  ${\bf k}=\omega{\bf n}/c$  is
the wave vector,  ${\bf n}$ is the wave front guiding vector, $\omega$
is   the   wave   frequency,   $k_4=i\omega/c$,   $k_1'   =(k_1+i\beta
k_4)/\gamma,   k_2'=k_2/\gamma,   k_3'=k_3/\gamma,   k_4'=k_4$,  ${\bf
k}'^2={\bf k}^2$ - inv.  (For comparison,  in the relativistic case we
have  $k_1'=(k_1+i\beta  k_4)/(1-\beta^2)^{1/2},  k_2'=k_2,  k_3'=k_3,
k_4'=(k_4-i\beta  k_1)/(1-\beta^2)^{1/2}$,   ${\bf   k}'^2+k_4'^2={\bf
k}^2+{k_4}^2$ - inv as is well-known).

The results obtained above we illustrate by means of the Table  1:  $$
\begin{array}{||c|c|c|c|c|c||}    \hline    \hline    Group    &P_{10}
&W_{11}&I&C_4 &G_1 \\ \hline WF \  \Phi(x)  &1  &\rho^l&x^2&\sigma(x)&
exp\{-i[(1-\gamma)k\cdot  x-\beta\omega(n_xt-x/c)]/\gamma\} \cr \hline
\hline \end{array} $$ For the  different  transformations  $x'=x'(x)$,
the weight functions $\Phi(x)$ may be found in a similar way.

Let us  note  that in the symmetry theory of D'Alembert equation,  the
conditions (\ref{f5})  for  transforming  this  equation  into  itself
combine the requirements formulated by various authors, as can be seen
in the Table 2:
$$\begin{array}{|l|l|l|l|l|} \hline \hline
Athor &Coordinates&Group&Conditions \ of \ invariance&Fields       \\
{}    &Transform.&{}   &{}                          &Transform.    \\
\hline
Voigt&x'_j=A_{jk}x_k&L_6X\Delta_1&A'_{ji}A'_{ki}=
{\rho'}^2\delta_{jk}&\phi'=\phi                                    \\
\lbrack4\rbrack&{}&{}&{}&{}                                        \\
Umov&x'_j={x_j}'(x)&W_{11}&
\displaystyle\frac{\partial x_j}{\partial  x'_i}\frac{\partial
x_k}{\partial          x'_i}={\rho}'^2\delta_{jk}&\phi'=\phi       \\
\lbrack12\rbrack&{}&{}&\Box' {x_j}=0&{}                            \\
Di Jorio&x'_j=L_{jk}x_k+&P_{10}&
L'_{ji}L'_{ki}=\delta_{jk}&\phi'=m_\alpha\phi_\alpha+              \\
\lbrack13\rbrack&a_j&{}&\displaystyle\frac{\partial^2\phi'}
{\partial\phi_\alpha\partial\phi_\beta}=0&m_0; \alpha=1,..,n       \\
Kotel'-&x'_j=x'_j(x)&C_4&\displaystyle
\frac{\partial      x_j}{\partial     x'_i}\frac{\partial
x_k}{\partial   x'_i}={\rho'}^2(x')\delta_{jk}&\phi'_\alpha=\psi
D_{\alpha\beta}\phi_\beta                                          \\
nikov&{}&{}&\displaystyle\frac{\partial^2\phi'_\alpha}{\partial
\xi_\beta\partial\xi_\gamma}=0&
\xi_\alpha=\psi\phi_\alpha                                         \\
\lbrack6-8\rbrack&{}&{}&\Box'\phi'_\alpha=0 \to &{}                \\
{}&{}&{}&\displaystyle \hat A\phi'_\alpha(\psi\phi_1,...\psi\phi_6)=0,
\Box\phi_\beta=0                       &\alpha,\beta=1,...,6       \\
{}&x'_j=x'_j(x)&G_1&\displaystyle\frac{\partial^2\phi'_\alpha}
{\partial\xi_\beta\partial\xi_\gamma}=0{}&\phi'_\alpha=\psi
M_{\alpha\beta}\phi_\beta                                          \\
{}&{}&{}&\Box'\phi'_\alpha=0 \to &\xi_\alpha=\psi\phi_\alpha       \\
{}&{}&{}&\displaystyle \hat B\phi'_\alpha(\psi\phi_1,...\psi\phi_6)=0,
\Box\phi_\beta=0                        &\alpha,\beta=1,...,6      \\
{}&{}&{}&{}&{}
\cr \hline \hline \end{array}$$
Here $m_\alpha,    m_0$   are   some   numbers, $D_{\alpha\beta}$   and
$M_{\alpha\beta}$ are the $6{\rm X}6$ numerical matrices.

According to this Table for  the  field  $\phi'=\phi$  with  conformal
dimension  $l=0$ and the linear homogeneous coordinate transformations
from   the    group    $L_6{\rm    X}\triangle_1\in    W_{11}$    with
$\rho=(1-\beta^2)^{1/2}$,  the  formulae were proposed by Voigt (1887)
[4, 9]. In the plain waves case they correspond to the transformations
of  the  4-vector  $k=({\bf k},  k_4)$ and proper frequency $\omega_0$
according to  the  law  $k_1'=(k_1+i\beta  k_4)/\rho(1-\beta^2)^{1/2},
k_2'=k_2/\rho,             k_3'=k_3/\rho,             k_4'=(k_4-i\beta
k_1)/\rho(1-\beta^2)^{1/2}$,  $\omega_0'=\omega_0/\rho$,  $k'x'=kx$  -
inv.  In the case of the $W_{11}$-coordinate transformations belonging
to the set of arbitrary transformations  $x'=x'(x)$  the  requirements
for the one component field with $l=0$ were found by Umov (1910) [12].
The requirement that the second  derivative  $\partial^2\phi'/\partial
\phi_\alpha\partial\phi_\beta=0$ with $\Phi=1$ be turned into zero was
introduced by Di Jorio (1974). The weight function $\Phi\ne 1$ and the
set  (\ref{f5}) were proposed by the author of the present work (1982,
1985, 1995) [6 - 8].

By now well-studied have been only the D'Alembert equation  symmetries
corresponding   to   the   linear  systems  of  the  type  (\ref{f7}),
(\ref{fg8}),  (\ref{f8}).  These are the well-known  relativistic  and
conformal  symmetry of the equation.  The investigations corresponding
to the linear conditions (\ref{f5}) are much more scanty and presented
only  in  the  papers  [6 - 8].  The publications corresponding to the
non-linear  conditions   (\ref{f2})   are   absent   completely.   The
difficulties arising here are connected with analysis of compatibility
of the set (\ref{f2}) containing the non-linear  partial  differential
equation.

Thus it  is  shown  that  with  the  generalized  understanding of the
symmetry according  to  Definition  1,  D'Alembert  equation  for  one
component  field is invariant with respect to any arbitrary reversible
coordinate transformations $x'=x'(x)$. In particular, they contain the
transformations of the conformal and Galilei groups realizing the type
of  standard  and  generalized   symmetry   for   $\Phi(x)=\phi'(x'\to
x)/\phi(x)$. The concept of partial differential equations symmetry is
conventional.

      \setlength{\parindent}{0mm}
\vspace{1mm}
{\bf     References}
\begin{list}{}{\setlength{\topsep}{0mm}\setlength{\itemsep}{0mm}
\setlength{\parsep}{0mm}}
\item[1.] N.N. Bogoliubov, D.V. Shirkov.  Introduction  in  Theory  of
Quantized Fields. Moscow, Nauka, 1973.
\item[2.] Yu.  Shirokov,  N.P.  Yudin. Nuclear Physics. Moscow, Nauka,
1972.
\item[3.] \begin{sloppypar}N.X.  Ibragimov.  Groups of Transformations
in Mathematical Physics. Moscow, Nauka, 1983.\end{sloppypar}
\item[4.] W.  Voigt.  Nachr. K. Gesel. Wiss., G\"ottingen, {\bf 2}, 41
(1887).
\item[5.] I.A. Malkin, V.I. Man'ko. JETP Lett., {\bf 2}, 230 (1965).
\item[6.] G.A.  Kotel'nikov.  Proc. Second Zvenigorod Seminar on Group
Theoretical Methods in Physics, V. 1. Ed. by M.A. Markov, V.I. Man'ko,
A.E. Shabad. (Chur, London, Paris, New York, 1985) 521.
\item[7.] G.A.  Kotel'nikov.  Proc.  Third   Yurmala   Seminar   Group
Theoretical  Methods in Physics.,  V.  1.  Ed.  by M.A.  Markov,  V.I.
Man'ko,  V.V. Dodonov. (Moscow, 1986) 479; Izv.VUZov, Fizika, {\bf 5},
127 (1989).
\item[8.] G.A.  Kotel'nikov.  Proc.   VII   International   Conference
Symmetry  Methods  in Physics,  V.  2.  Ed.  by A.N.  Sissakian,  G.S.
Pogosyan. (Dubna, 1996) 358; http://arXiv.org/abs/physics/9701006
\item[9.] W.   Pauli.   Theory   of   Relativity.    Moscow-Leningrad,
Gostexizdat, 1947.
\item[10.] W.I.  Fushchich,  A.G.  Nikitin.  Symmetries   of   Quantum
Mechanics Equations. Moscow, Nauka, 1990.
\item[11.] P.  Carruthers. Phys. Reports (Sec. C of Phys. Lett.), {\bf
1}, 2 (1971).
\item[12.] N.A.  Umov. Collected Works, Moscow-Leningrad, Gostexizdat,
1950.
\item[13.] M.  Di Jorio. Nuovo Cim., {\bf 22B}, 70 (1974).
\end{list}
\end{document}